# Detecting Nonlinearity in Data
# with Long Coherence Times

**James Theiler**

*Center for Nonlinear Studies and Theoretical Division,*
*Los Alamos National Laboratory, Los Alamos, NM 87545;* and
*Santa Fe Institute, 1660 Old Pecos Trail, Santa Fe, NM 87501*

**Paul S. Linsay**

*Plasma Fusion Center, Massachusetts Institute of Technology,*
*175 Albany Street, Cambridge, MA 02139*

**David M. Rubin**

*U. S. Geological Survey, 345 Middlefield Road, Menlo Park, CA 94025*

(Friday, November 13, 1992)

We consider the limitations of two techniques for detecting nonlinearity in time series. The first technique compares the original time series to an ensemble of surrogate time series that are constructed to mimic the linear properties of the original. The second technique compares the forecasting error of linear and nonlinear predictors. Both techniques are found to be problematic when the data has a long coherence time; they tend to indicate nonlinearity even for linear time series. We investigate the causes of these difficulties both analytically and with numerical experiments on "real" and computer-generated data.

In particular, although we do see some initial evidence for nonlinear structure in the SFI dataset E, we are inclined to dismiss this evidence as an artifact of the long coherence time.

*"May you have interesting data."*
— ancient Chinese curse

# 1 Introduction

For time series that arise from chaotic systems, there are certain quantities (*e.g.*, the fractal dimension of the strange attractor, or the spectrum of Lyapunov exponents) that are especially interesting because they characterize intuitively useful concepts (number of active degrees of freedom, or rate of divergence of nearby trajectories) and are invariant to smooth coordinate changes. Algorithms for estimating these quantities are available, but they are notoriously unreliable, and often rely heavily on the skill and judgement of their operators. There is an embarrassing lack of consensus, even among the so-called experts, on what constitutes a good estimate of dimension or Lyapunov exponent, or even whether chaos is present in a give time series. To some extent this difficulty may be attributed to inadequate comparison of one algorithm to another (and this conference is aimed at addressing that inadequacy), but to some extent, it is just a hard problem.

The problem is arguably hard enough for long noise-free data sets generated on a computer from low-dimensional maps or differential equations. For "real" data, as the speakers at this conference have repeatedly emphasized, the problem is far more difficult. (And as the organizers have repeatedly reminded us, far more valuable.) Real data is contaminated with noise (which is rarely additive, Gaussian, or white), is measured with finite precision, and is subject to innumerable external influences in the environment and the measurement apparatus. And of course there is never enough of it.

In this article, we will describe (yet) another source of difficulty that arises in the analysis of time series data. The particular problem of detecting nonlinear structure — either by comparison of the data to linear surrogate data, or by comparing linear and nonlinear predictors — is seen to be complicated when the data exhibits long coherence times.

In this section we define some terms and discuss linear modeling of time series. Section 2 describes the method of surrogate data, and compares two approaches to generating surrogate data. We find that both have difficulties trying to mimic data with long coherence time. We illustrate these problems with real and computer-generated time series in Section 3, including the time series E.dat from the the SFI competition. In the last section, we discuss what it is about the analysis or the data that is problematic.

## 1.1 Terminology

A *time series* is a sequence of measurements $x_1, x_2, \ldots, x_N$ of some physical system taken at regular intervals of time. A time series can be thought of as a particular realization of a stochastic *process*, which we will define as a sequence of random[1] variables $\ldots, X_{-1}, X_0, X_1, X_2, \ldots$. We make this distinction because theorems and formal definitions are available only for *processes*, while the whole purpose of generating this formalism is to assist researchers who are confronted with real experimental *time series*.

---

[1]Note that even a deterministic process is usefully defined as a sequence of random variables. For the logistic process, $X_{t+1} = 4X_t(1 - X_t)$, for instance, each variable $X_t$ has a nontrivial probability distribution $P(X_t)$, but the joint distribution $P(X_{t+1}, X_t)$ reflects the deterministic law.

We will also distinguish the terms *system* and *model*, by letting *system* refer to the actual underlying physics,[2] and *model* to a hypothetical description of the system. Since the *model* will (for our present purposes) be inferred only from the time series, we cannot expect the model to be expressed in terms of the physics. But, although the model is really nothing more than an operational description of the time series, the hope is that this description — in conjunction with knowledge of the appropriate physics — will actually say something useful about the underlying physical system. When we talk about a *best* or "correct" (always in quotation marks!) model, we will mean the model which — out of a (usually parametric) family of models — has the least root mean squared (rms) error in its one-step-ahead forecast.[3]

Three statistics of particular interest are the *mean* $\mu = \langle X_t \rangle$, the *variance* $\sigma^2 = \langle (X_t - \mu)^2 \rangle$, and the *autocorrelation function* $A(\tau) = \langle (X_t - \mu)(X_{t-\tau} - \mu) \rangle / \sigma^2$. Here $\langle \cdot \rangle$ represents, for the process, an ensemble average. If the process is ergodic, the average could also be over time $t$, and in that case, good *sample* statistics can be defined from a single *time series*.

When we speak of "coherence time," what we mean is the time beyond which a signal becomes uncorrelated with its past. We can formalize the concept somewhat by defining coherence time as that time $\tau$ such that the absolute value of the autocorrelation function $|A(T)|$ is smaller than some pre-specified value $\epsilon$ for all $T > \tau$. This is to be distinguished from the *first* time $T$ that the autocorrelation $A(T)$ drops to a value below $\epsilon$; in other words, we are interested in the "envelope" of the autocorrelation curve. This is not really satisfactory as a formal definition of coherence time — for one thing, it depends on the choice of $\epsilon$ — but it is adequate for our current purposes.

In general, however, if the autocorrelation $A(T)$ vanishes exponentially fast as $T \to \infty$, we will say that the coherence time is finite; if it does not vanish at all (if $\limsup_{T \to \infty} |A(T)| > 0$), then we say that the coherence time is infinite.[4]

## 1.2 Wold decomposition

The Wold decomposition is the fundamental theorem of linear time series analysis (*e.g.*, see Ref. [2, §7.6.3]). This theorem states that any stationary zero-mean process (linear or nonlinear) can be decomposed into the sum of two uncorrelated components: one "deterministic" and one "indeterministic." That is

$$x_t = z_t + u_t \tag{1}$$

where the linearly deterministic $z_t$ can be modeled exactly with a (possibly infinite) linear combination of past values, and where the indeterministic $u_t$ can be modeled by a moving

---

[2]Mathematically, the *system* is equivalent to the *process*, but the connotation we mean to imply for a *system* is that it is physical.

[3]This is a convenient but basically arbitrary criterion. As Tsay [56] emphasizes, "it is well known that the best model with respect to one checking criterion may fare badly with respect to another criterion."

[4]One definition that is consistent with these constraints is $\tau = \sum_{T=1}^{\infty} \sqrt{1 - E^2(T)}$, where $E(T)$ is the rms forecasting error $T$ time steps into the future for the best linear model, normalized by the standard deviation of the data. We won't actually be using this definition (the informal description in the text will be adequate), but it does seem appropriate to at least write such a definition down.

average of uncorrelated innovations.[5]

$$z_t \;=\; \sum_{i=1}^{\infty} \alpha_i z_{t-i} \tag{2}$$

$$u_t \;=\; \sum_{i=0}^{\infty} \beta_i e_{t-i} \tag{3}$$

It can be shown that the autocorrelation $A_z(T) = \langle z_t z_{t-T}\rangle / \langle z^2\rangle$ of the deterministic component of the time series will be significantly nonzero for arbitrarily large $T$, whereas the autocorrelation $A_u(T)$ of the indeterministic time series will approach zero as $T$ becomes large. Since $u$ and $z$ are uncorrelated (again, not necessarily independent, but satisfying $\langle u_t z_t\rangle = 0$), it follows that the autocorrelation in the full time series is

$$A_x(T) = \frac{A_u(T)\langle u^2\rangle + A_z(T)\langle z^2\rangle}{\langle u^2\rangle + \langle z^2\rangle}. \tag{4}$$

From the point of view of the Wold decomposition, then, a process has a finite (resp. infinite) coherence time if and only if its linearly deterministic component is zero (resp. nonzero).

## 1.3  Linear modeling of time series (ARMA)

It follows from the Wold decomposition theorem that any stationary process can be modeled as an autoregressive moving-average:

$$x_t = x_o + \sum_{i=1}^{\infty} a_i x_{t-i} + \sum_{i=0}^{\infty} b_i e_{t-i}. \tag{5}$$

For instance, given $\alpha_i$ and $\beta_i$ from Eqs. (2-3), one can take $a_i = \alpha_i$ and $b_i = \beta_i - \sum_{j=1}^{i} \alpha_j \beta_{i-j}$. However, this is not necessarily a unique solution. For indeterministic time series, for instance, it is possible to write the time series as a pure auto-regressive (AR)

$$x_t = x_o + \sum_{i=1}^{\infty} a_i x_{t-i} + \sigma e_t, \tag{6}$$

or as a pure moving-average (MA)

$$x_t = x_o + \sum_{i=0}^{\infty} b_i e_{t-i}. \tag{7}$$

For time series with infinite coherence time (nonzero linearly deterministic component), however, a full ARMA model is typically required.

In the study of linear Gaussian processes, the innovations are taken to be independent and identically distributed (IID) Gaussian random variables. For indeterministic time series, which can be written as a pure moving average of the Gaussian innovations, this implies that the time series itself will be Gaussian. However, if a deterministic component is present (again, that means an infinite coherence time) then Gaussian innovations do not necessarily imply Gaussian data. For example a sine wave with added Gaussian white noise can be modeled as a linear process with Gaussian innovations but the time series is not Gaussian.

---

[5]These innovations are uncorrelated, or "white," but they are not necessarily independent. This means $\langle e_t e_{t'}\rangle = 0$ for $t \neq t'$, but *not* that the joint distribution $P(e_t, e_{t'})$ is equal to the product of the marginal distributions $P(e_t)P(e_{t'})$. The innovations are treated as "noise" in linear analysis, but they may well possess nonlinear deterministic structure. The Wold decomposition is quite general, and applies to all stationary processes, including low-dimensional chaos.

Lii and Rosenblatt [31] have discussed linear (indeterministic) processes with non-Gaussian innovation; they show that these processes are far more complicated than those with Gaussian innovations.

## 2 Surrogate data

Surrogate data is artificially generated data which is to be used in place of an original data set; the main purpose is to provide a kind of baseline or control against which the original data can be compared. In tests for chaos, for example, one can control against artifacts due to autocorrelation in a time series by generating surrogate data from a random process that mimics the autocorrelation of the original time series. Suppose some algorithm indicates low-dimensional chaos in a time series. If the same algorithm also indicates low-dimensional chaos in the surrogate time series, then one can dismiss the original evidence for chaos as an artifact of the autocorrelation.

More formally, the method provides a mechanism for testing well formulated null hypotheses. It can be difficult to precisely formulate *interesting* null hypotheses, and often very difficult to prescribe a surrogate data generator which is appropriate for such a null hypothesis. Our work has focused on tests for nonlinearity which take linearly correlated Gaussian noise as the null hypothesis. In this case, one is not looking for chaos *per se*, but for some statistic which is significantly different for the original time series than it is for the linear surrogates. The existence of such a statistic implies that the original time series is inconsistent with the null hypothesis, and therefore that the original time series is nonlinear.

While the systematic application of this approach to tests of potentially chaotic time series has only recently become fashionable, the basic idea is by no means new. Monte-Carlo methods for generating data sets with specified properties are widely used, and in some applications have reached the status of recipes [38, §14.5]. Statisticians have long advocated resampling (so-called "bootstrap") methods, in which new data sets are generated by randomizing the original data set in some prescribed way. We have found the writing of Efron in particular to be enlightening and inspirational [10, 11]. The purpose of these methods, however, is usually not to test a hypothesis, but to estimate confidence intervals for some statistic of interest.

The application of these resampling methods to time series is complicated by the temporal dependence of time series data; most of the original bootstrap applications considered individual data points to be independent events. An indirect approach is to remove the linear dependence in the data by considering the innovations (the "residual" time series) of an ARMA model [11], though the filtering required to produce the residuals can make it harder to "see" the nonlinearity in a chaotic time series [50]. Direct resampling techniques based on temporal "blocks" of data were discussed by Künsch [26], and an improvement was developed by Politis and Romano [30, 37]. While further exploration is certainly called for, it is not clear to us that these methods (at least as they have been applied in Refs. [26, 30, 37]) can be used in conjunction with dynamical statistics for the purpose of hypothesis testing.

Parametric bootstraps, instead of resampling the data directly, use the data to set parameter values, and then use these values in a parametric model for generating new data. An incomplete list of authors who have successfully used this approach include: Grassberger [15], who used a simple linear autoregressive process to generate a time series which mimicked properties of a climate data set originally purported to exhibit low-dimensional chaos; Kurths and Herzel [27], who compared estimates of dimension and Lyapunov exponent for a time series of solar radio pulsations with those for data from an AR(5) model that fairly accurately matched the spectral properties of the original data; Brock *et al.* [5], who generated surrogate financial time series to test trading strategies; Ellner [12], who used this approach to show that a variety of nonchaotic "plausible alternatives" might adequately explain measles and chickenpox data; and Tsay [56], who provides an excellent overview of the approach with a wide variety of applications.

Kaplan and Cohen [23] published the first example we are aware of in which the evidence for chaos in a time series was evaluated by comparing against a control data set that was generated by the Fourier transform (FT) method which is described in Section 2.1. Somewhat earlier, Osborne *et al.* [35, 36] inverted $1/f^\alpha$ spectra using an inverse Fourier transform to generate realizations of $1/f^\alpha$ noise, and then showed that dimension estimates of these time series were problematic. (This issue has been further discussed in Refs. [39, 49].) The use of multiple surrogate data sets for more formal statistical hypothesis testing was suggested in Refs. [47, 48] and implemented in Refs. [51, 52] for a variety of examples. Smith has applied the surrogate data methods to fluid dynamical time series [45], and more recently, to address the issue of inherent periodicities in the climate record.[6] A variant of the surrogate data approach has also been described in Ref. [25].

The use of formal statistics, in which the null hypothesis is explicitly spelled out and carefully tested against, is only lately gaining popularity in the chaos community. Brock, Dechert, and Scheinkman [4] deserve to be singled out for creating perhaps the first statistically rigorous application of the Grassberger-Procaccia [16] correlation integral for time series analysis. This work has led to a veritable industry in the economics community involving the application of statistics which incorporate the explicit recognition of chaos [3, 6, 7, 19, 20, 28, 29, 43]; these complement the more classical approaches taken by the statisticians [18, 24, 33, 46, 54–56]. Many of these are reviewed in Tong's comprehensive book [53].

## 2.1 FT-based surrogates

To test for nonlinearity, we begin with the pre-supposition that the time series is linear. A more precise formulation of the null hypothesis is that the data arise from a linear stochastic process with Gaussian innovations.[7]

The algorithm we generally use for making linear surrogate data is based on the Fourier transform (FT). Specifically, we compute a discrete Fourier transform of the original data, and replace the phases at each frequency with random numbers in the interval $[0, 2\pi)$ while keeping the magnitude at each frequency (*i.e.*, the power spectrum) intact[8], and then apply

---

[6]L. A. Smith (personal communication).

[7]An extended null hypothesis which considers that there is an underlying process that is Gaussian, but one is observing a static nonlinear transform of that process, is discussed in Ref. [51].

[8]It is important that the phases be symmetrized in such a way that the inverse Fourier transform is real

the inverse Fourier transform to produce the surrogate time series.

This is a kind of nonparametric bootstrap which by construction produces surrogates that have the same power spectrum as the original data. In fact, the surrogate time series have exactly the same *sample* power spectrum as the original time series. The Wiener-Khintchine relations assure us that two processes with the same power spectrum will also have the same autocorrelation function, but in comparing the sample statistics, we have to be more careful. Jenkins and Watts [21] note that there are (at least) three different ways to define a sample autocorrelation (In these definitions, the time series is for convenience assumed to have zero mean):

- Unbiased estimator: $\frac{1}{N-T} \sum_{t=1}^{N-T} x_t x_{t+T}$.

- Biased estimator (lower variance than unbiased estimator): $\frac{1}{N} \sum_{t=1}^{N-T} x_t x_{t+T}$.

- Circular autocorrelation: $\frac{1}{N} \left( \sum_{t=1}^{N-T} x_t x_{t+1} + \sum_{t=N-T+1}^{N} x_t x_{t+T-N} \right)$.

The estimators agree to order $O(T/N)$, and for $T \ll N$ and $N \to \infty$ all three approach the actual autocorrelation of the process. But for finite $N$ they are only approximately equal. And of the three, it is the circular autocorrelation that is exactly preserved in going from the original to the surrogate data sets.

For a Gaussian linear *process*, all of its properties are encoded in the mean, variance, and autocorrelation. But when we say that the "linear" properties of the *time series* are preserved in the surrogate time series, what that means exactly is that the sample mean, sample variance, and circular autocorrelation are preserved.

## 2.2 ARMA model-based surrogates

Instead of attempting to exactly preserve some preselected set of sample statistics, an alternative approach for generating surrogate data is to directly fit the data to a constructive parametric linear model, such as a finite-order ARMA$(p, q)$:

$$x_t = x_o + \sum_{i=1}^{p} a_i x_{t-i} + \sum_{i=0}^{q} b_i e_{t-i}. \tag{8}$$

Constructing a parametric model from a finite set of data involves choosing the "correct" values for $q$ and $p$, and this is an issue of some subtlety; one wants enough terms to capture the correct correlations in the data but not so many terms that the data is over-fit. Aikake [1] and Schwarz [44] have suggested fairly general criteria; a more recent discussion specific to the ARMA model can be found in Ref. [40]. For fixed values of $q$ and $p$, the optimal parameters $(a_i, b_i)$ depend in principle only on the autocorrelation of the stochastic process.

If there is no deterministic component ($z_t = 0$ in Eq. (1)), then an ARMA$(p, q)$ process can be modeled by a pure autoregressive AR model or a pure moving-average MA model, of appropriately large order.[9] We note that in practice it is much easier to fit coefficients

---

and the power at each frequency is unaffected; we remark that the recipe for doing this in Ref. [52, §A.1,#4] is incorrect. We are indebted to W. Schaffer for pointing out this error.

[9]But in general a pure AR or pure MA will be less parsimonious than the best ARMA$(p, q)$ model; that is, the AR or MA models will usually require more than $p + q$ parameters. Having said this, we should further note that the ARMA formalism doesn't necessarily generate the most parsimonious description of linear Gaussian processes either.

to a pure AR model than to an MA or ARMA model. In that case, assuming a zero-mean process for convenience, the formula is given by [2, p. 187]:

$$
\begin{bmatrix}
1 & A(1) & \cdots & A(m-1) \\
A(1) & 1 & \cdots & A(m-2) \\
\vdots & \vdots & \ddots & \vdots \\
A(m-1) & A(m-2) & \cdots & 1
\end{bmatrix}
\begin{bmatrix}
a_1 \\
a_2 \\
\vdots \\
a_m
\end{bmatrix}
=
\begin{bmatrix}
A(1) \\
A(2) \\
\vdots \\
A(m)
\end{bmatrix}.
\tag{9}
$$

These are sometimes called the Yule-Walker equations.

Having determined the appropriate ARMA model, one can generate surrogate data by inserting Gaussian IID random numbers into the $e_t$ terms, and then iterating Eq. (8). One is assured that in the long run, the autocorrelation in the surrogate data will approach the autocorrelations used in Eq. (9), but note that this is different from the exact match of sample statistics that is seen in the FT surrogates.

A common alternative practice is is to bootstrap the residuals themselves. Having fit the model to the data, one derives a time series of residuals $e_t$ which are then scrambled and re-inserted into Eq. (8). This avoids the assumption of Gaussian innovations, and therefore leads to a broader class of time series, and presumably tests against a looser null hypothesis; however, linear processes with non-Gaussian innovations do not always behave in "linear" ways — for instance, see Tong [53, pp. 13-14],Lii and Rosenblatt [31], or Kanter [22] for examples of some of the pathologies.

It is also worth noting that this AR model is also the optimal linear predictor; that is the average squared errors

$$
\langle E_t^2 \rangle = \left\langle \left( X_t - \left[ x_o + \sum_{i=1}^{m} a_i X_{t-i} \right] \right)^2 \right\rangle
\tag{10}
$$

are minimized when the coefficients $a_i$ are chosen according to Eq. (9).

## 2.3 Comparison of FT and ARMA surrogates

The superficial equivalence of FT and ARMA modeling rests with the notion that both the Fourier spectrum and the AR coefficients depend only on the autocorrelation function of the original time series, which is (at least approximately) mimicked by the surrogates in both cases.

The difference between FT and ARMA surrogates is basically the difference between "fitting the data" versus "fitting the model." FT surrogates exactly match certain sample statistics (mean, variance, and circular autocorrelation) of the original data. ARMA surrogates are generated from a model that is *fit* to the original time series. These surrogates exhibit sample statistics that are usually but not necessarily in approximate agreement with those of the original time series.

Another difference is the way the data sets are generated. The FT method makes a whole new time series all at once, and it necessarily has the same length as the original time series. The ARMA method generates new points iteratively, one at a time, and can generate arbitrarily long or short data sets. This is not necessarily an advantage, though. Generating points sequentially, one is vulnerable to instabilities that may amplify small errors into large effects in the long term. This is a general difficulty with model-based surrogate data methods;

two models which are approximately equal (say, have nearly equal ARMA coefficients) can give rise to time series that are markedly different.[10] A second difficulty that arises when modeling processes with long coherence times is that the qualitative long term behavior, even the overall amplitude of the process, can depend not only on the model parameters but on initial conditions as well.

While the FT method is nonparametric, in the sense that one does not directly fit a model to the data, one can think of it rather as having a very large number of parameters, $N/2$, corresponding to the amplitude of the power spectrum at $N/2$ frequencies. As a model, then, the FT provides an extreme overfit to the data. By contrast, ARMA models are parsimonious, in that the modeler is (usually) careful to choose the minimum number of parameters needed to fit the data.[11]

Which approach is preferable depends on the application. Our view is that FT surrogates are better for testing hypotheses, while ARMA surrogates may be better for estimating confidence intervals. Certainly the FT surrogates will be useless for estimating confidence intervals for estimates of mean, variance, and autocorrelation.

## 3   Application to time series data

In this section, we will investigate four different time series. The first is a real time series that was part of the SFI competition. Though we seem to see evidence for nonlinearity in this time series, we give reasons to suspect the results. The second data set is an artificially generated sine wave plus noise. This data is meant to be a caricature of the real data, but a caricature whose underlying process is known. With this second data set we are able to see the same effects that we observed with the "real" data set, and thereby confirm our suspicions that the effects we saw were artifacts of the long coherence time. The third data set is also, strictly speaking, a sine wave plus noise, but it is a particularly simple example that permits some analytical discussion. For the third data set, we compare the theoretical efficiency of linear versus nonlinear predictors. Finally the last data set is a sum of two commensurate sine waves with some added noise; in this case, we see numerically we we described in theory for the third data set: namely, that the prediction error of a nonlinear model fit to the data is smaller than the error of a linear model fit to the data.

### 3.1   The investigation of E.dat

We apply tests for nonlinearity based on the method of surrogate data to the SFI data set E.dat. These data are observations of the light curve of a variable white dwarf star, and are sampled every ten seconds. We concentrate on a single series #14, chosen more or less arbitrarily.[12] Fig. 1(b) shows the first $N = 2048$ points of this time series. The most noticeable feature is the coming and going of an oscillation with a period of 50 time units (500 seconds, or about 8.3 minutes). We computed discrete Fourier transforms on all all seventeen data sets, using data segments of varying length and location in the time series, and both with and without a Hanning window. (See Fig. 2(a) for a particular case.) We see considerable variation, and would not be confident in attempting our own detailed interpretation of the power spectrum.[13] However, we do consistently see two peaks in the vicinity of the dominant frequency (0.002 Hz), suggesting that the signal is quasiperiodic and that the "coming and going" may be a beating phenomenon. The autocorrelation curve (Fig. 2(b)) supports this interpretation, and also indicates that the coherence time is at least on the order of a thousand time steps, and possibly much longer.

---

[10]For nonlinear modeling, this can be extremely problematic. A parametric model that exhibits chaotic behavior, for instance, can with an arbitrarily small change in parameter, give rise to stable periodic behavior. This is sometimes referred to as the genotype/phenotype conundrum. One associates genotype with equations of motion, and phenotype with the long term behavior of those equations. Small perturbations in the genotype can give rise to huge differences in phenotype. And inferring the genotype from the phenotype is much more difficult than the other way around. We should remark that for linear modeling, the difficulty is not this extreme. In this case, if the roots of the characteristic polynomial of the AR part of the model are well within the unit circle, then a small perturbation of parameters will not grossly affect the overall behavior. (However, we might also remark that for high order polynomials, small changes in the coefficients can lead to large changes in the roots.)

[11]The problem of parsimony and "effective number of parameters" in nonlinear modeling is much more subtle in the case of nonlinear modeling; see Refs. [34, 57] for interesting discussions of this issue.

In searching for nonlinear structure, any nonlinear statistic in principle is adequate. We used an estimator of fractal dimension, obtained from the slope of a correlation integral [17] at a point $r$ equal to half of the rms amplitude of the time series. While this is not our best shot at what the actual dimension is (in fact, for this data, we do not really even see a hint of low-dimensionality), it does provide a nonlinear statistic against which we can compare real data to surrogate. What we see in Fig. 3(a) is that — for this statistic — the real and surrogate data are indistinguishable. We quantify significance by counting the number of "sigmas" between the original and surrogate values for the discriminating statistic, where a "sigma" is the standard deviation of all the values of the statistic computed for the surrogate data sets.

Because the data set has a lot of what appears to be high frequency noise, we also considered a crude low-pass linear filter of the data, based on a moving average (equal coefficients) of ten sample points. That is, $x'_t = (x_t + x_{t-1} + \cdots + x_{t-9})/10$. Fig. 1(e) shows how smoothing affects this data set, and in Fig. 3(b), we again compare real data to surrogates. At about the four sigma level, the difference between the real data and surrogates is statistically significant. Inspection of the actual values, however, reveals that the difference is never more than 8%; we are inclined to remark that the difference is "significant," but not very "substantial." When we used nonlinear forecasting error instead of estimated dimension as our discriminating statistic, we did not see any significant evidence for nonlinearity for either the smoothed or the raw data set.

Now, if the surrogate data really is mimicking all the linear properties of the original time series, then any linear statistic computed from both surrogate and original data should give the same value. We plot one such statistic, the in-sample fit error of the best linear model, in Fig. 4(a,b). For both E14.dat and the smoothed E14.dat, there is a small but statistically significant discrepancy. So the surrogate data evidently is *not* mimicking "all" of the linear properties. The technical explanation is that the in-sample fit error is a sample statistic which does not depend precisely and entirely on the circular autocorrelation. That the discrepancy should be systematic, however, is an artifact of long coherence times, as we show in Section 3.2.

---

[12]We were partly motivated to use this series because we knew that M. Paluš (in this volume) had looked at the same series.

[13]We have not attempted to use the information (which was provided) which gave the absolute starting times for each of the seventeen time series. Combining the data into one long time series with appropriate gaps should permit much more precise spectral estimation.

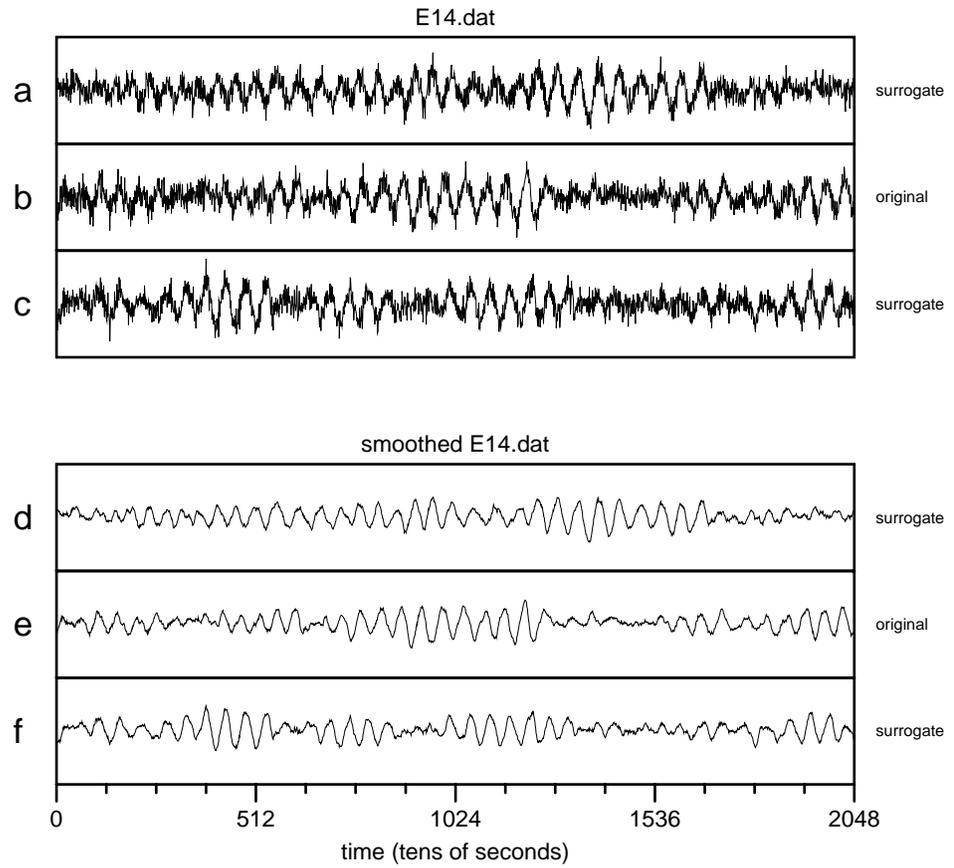

Figure 1: (a,b,c) The top three time series are (b) SFI data set E.dat #14, and (a,c) two surrogate data sets. (d,e,f) The bottom three are (e) set E.dat #14 smoothed with a moving average window of size ten time steps, and (d,f) two of its surrogates. Figures (b,e) are the first $N = 2048$ points of an approximately 2600 point data set.

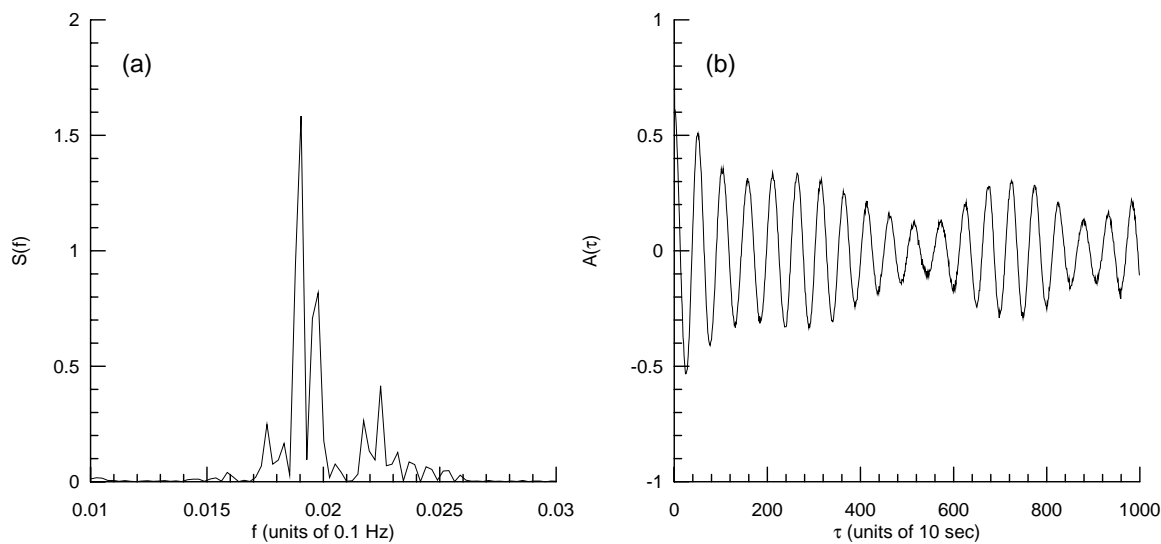

Figure 2: *(a) Power spectrum, computed by a discrete Fourier transform from all N = 2602 points in the time series E14.dat, using a Hanning window. (b) Autocorrelation of the data set E14.dat computed with the biased estimator $A(T) = \frac{1}{N}\sum_{t=1}^{N-T} x'_t x'_{t+T}$, where $x'_t = (x_t - \mu)/\sigma$ is the normalized time series value.*

## 3.2  Sine wave plus noise

To investigate the effects of long coherence time in a situation where we know the underlying process, we generated artificial data with infinite coherence time by adding measurement noise to an underlying sine wave. We chose the period and noise level to (very crudely) approximate that of the smoothed E.dat.

In general, as Fig. 5 shows, generating surrogate data by the FT algorithm leads to surrogates that do not have the coherent structure of the original sine wave. It *is* possible to generate good surrogates by fortuitous choice of data length. For periodic data, this is only a slight inconvenience (requiring the use of a general discrete Fourier transform (DFT) instead of the fast Fourier transform (FFT) which requires data length to be a power of two): for quasiperiodic data, this is trickier, because one must choose the length of the time series to be (at least approximately) commensurate with *both* periods.

There are two effects going on here. The first involves choosing the length so that the periodic continuation is at least continuous (doesn't have a jump). If this is not done, one introduces spurious high frequencies into the data. This effect can be alleviated to some extent by windowing the data, *e.g.*, with a Hanning window (see Ref. [51, §2.4.2]). The second effect involves choosing the length of the time series so that *all* the relevant periods are commensurate with this length. If this is not done, then the DFT takes the power from a single frequency and distributes it to adjacent frequency bins; upon inverting the DFT after randomizing the phases, one sees a beating between the adjacent frequencies instead of the pure frequency in the original time series. In this second case, windowing the data does not help.

Using this sine wave plus noise, we see in Fig. 6 that a dimension-based test for non-linearity is able to distinguish the real and surrogate data with high statistical confidence. Again, the difference is extremely "significant" but not especially "substantial." Further, as Fig. 4(c) shows, the data is also distinguished from the surrogates by a linear statistic.

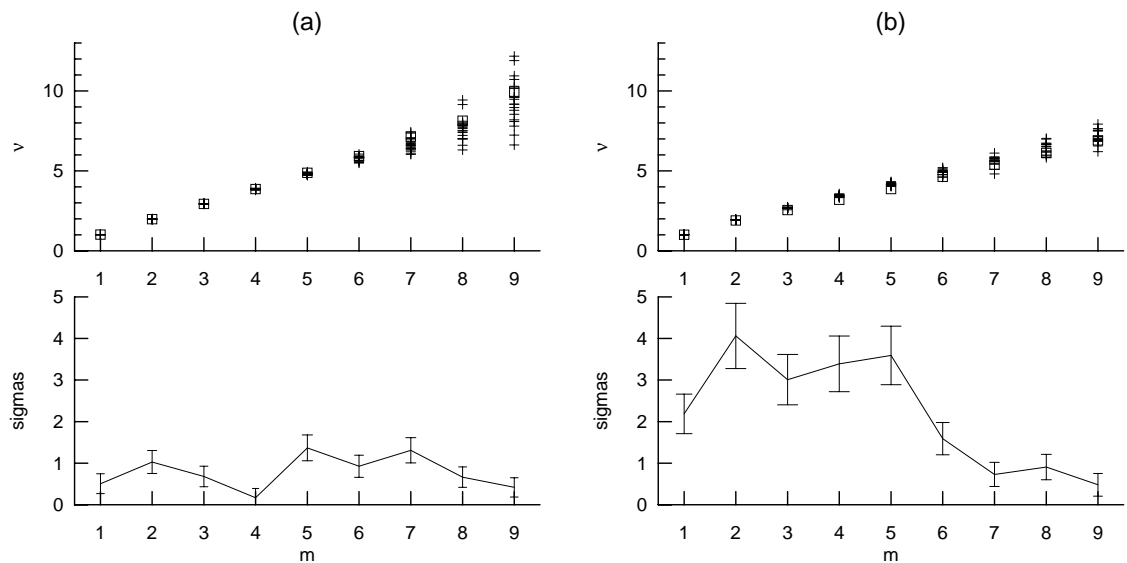

Figure 3: *Significance of evidence for nonlinearity based on an estimate of the correlation dimension: (a) for data set E14.dat, and (b) for the smoothed data set. The top panels show estimated dimension for real (□) and surrogate (+) data, as a function of embedding dimension. The bottom panels show "number of sigmas" as an indication of the statistical significance with which one can reject the null hypothesis that the experimental data is linear.*

We argued in Section 3.1 that the evidence for nonlinearity observed in E.dat might be an artifact of the long coherence time. In this section, we have shown that the effects seen with E.dat are also seen in a data set which by construction is linear, but which has a long (in fact, infinite) coherence time.

## 3.3  Linear versus nonlinear modeling: an example

Another way to test for nonlinearity in a time series is to compare the linear and nonlinear models to see which more accurately predicts the future. For example, Casdagli [8, 9] has described an "exploratory" approach in which the data is fit with local linear models using $k$ nearest neighbors. The parameter $k$ is swept from $m + 1$, the minimum value required to make a local linear fit in $m$ dimensions, up to the size $N$ of data set itself. For $k < N$, the model is nonlinear, but for $k = N$ it is equivalent to a globally linear model. If the error decreases monotonically with $k$, then the process is taken to be linear. If the error increases monotonically with $k$, then the process is taken to be nonlinear and deterministic. If, as most often happens, the error first decreases with increasing $k$ and then increases, the process is taken to be nonlinear and stochastic.

Kanter [22] has shown the unsurprising result that for an indeterministic linear Gaussian process, the optimum predictor is a linear predictor.[14] We consider a slightly different case than Kanter studied; we look at a *time series* that is generated by a linear *process*, and compare the linear and nonlinear models that are *fit* to the finite length of the time series. What we find for data with long coherence times is that the nonlinear models are often superior.

---

[14]What *is* surprising in Kanter's paper is that linear non-Gaussian processes *can* be more accurately modeled with a nonlinear predictor. We are grateful to J. Scargle for pointing out this reference to us.

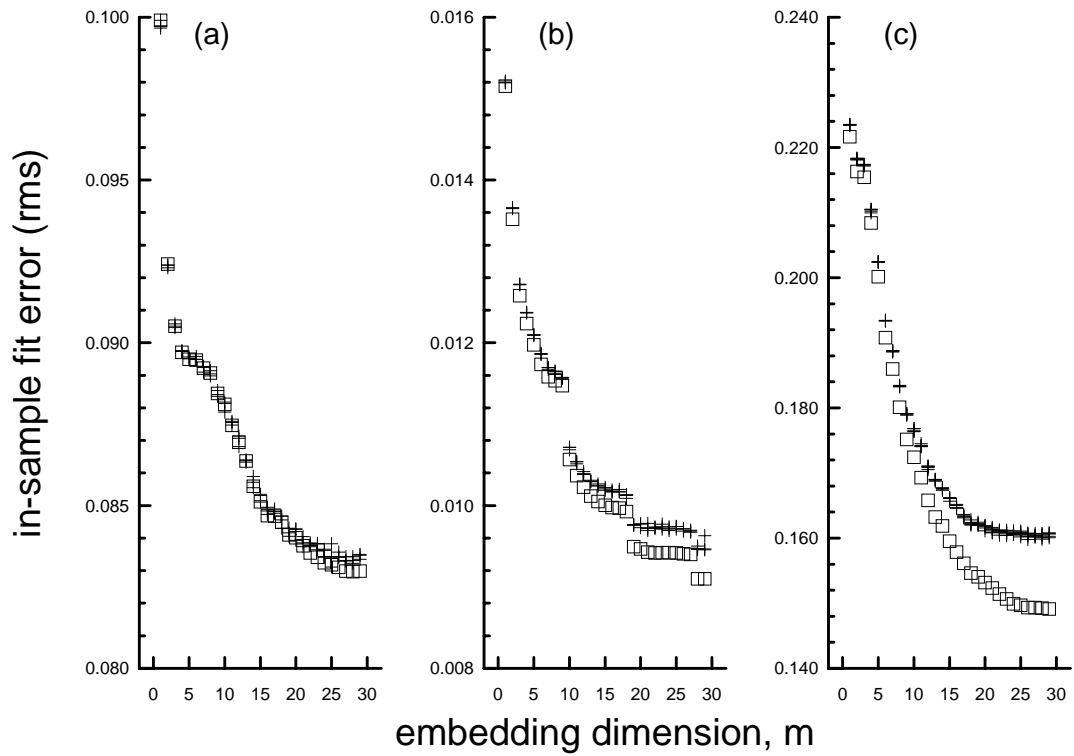

Figure 4: *A linear statistic, the in-sample rms fitting error, is computed for linear models with embedding dimension m. Here (a) is for E14.dat, (b) is for the smoothed data, and (c) is an artificially generated sine wave with measurement noise. It is apparent, particularly for cases (b) and (c), that the surrogate data is not as linearly predictable as the original data.*

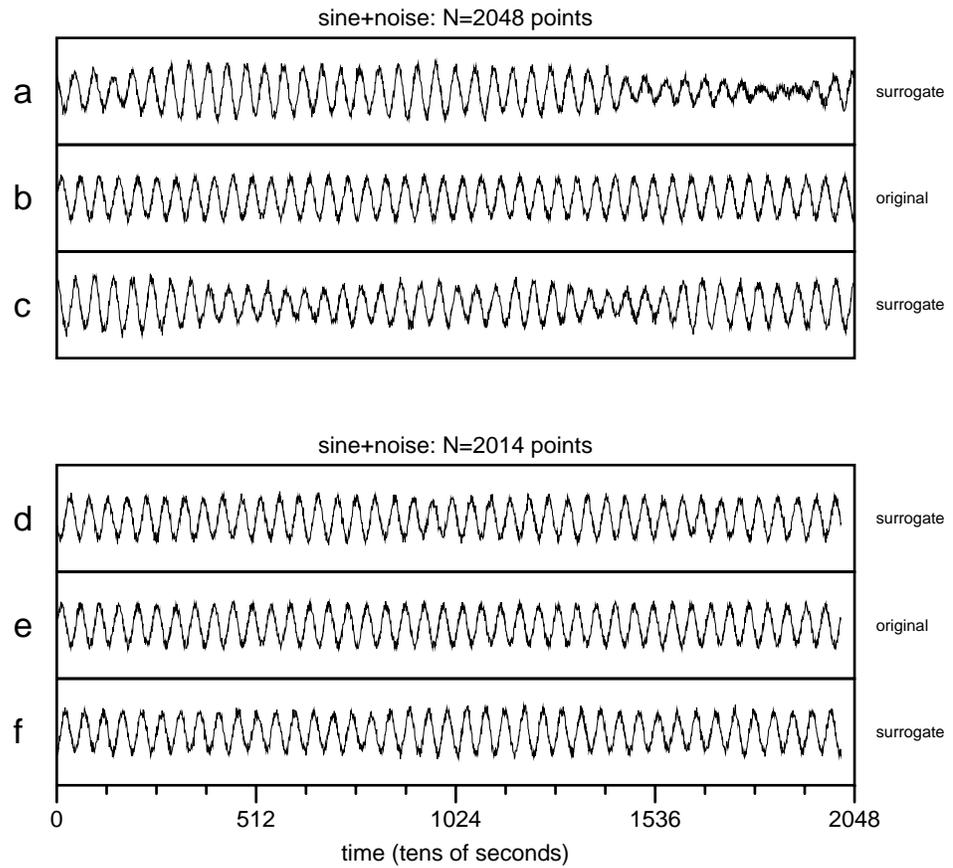

Figure 5: *Sine signal with white measurement noise, and surrogate data sets generated by the FT algorithm without windowing. In (a,b,c), the length of the time series is a convenient (for FFT purposes) power of two, N = 2048; the original data set is in (b), while (a) and (c) are the surrogate data sets. In (d,e,f), we use the same time series, slightly truncated to N = 2014 points, so that there is a near-integral number of oscillations in the time series. Again the middle data set (e) is the original, while (d) and (f) are two surrogates.*

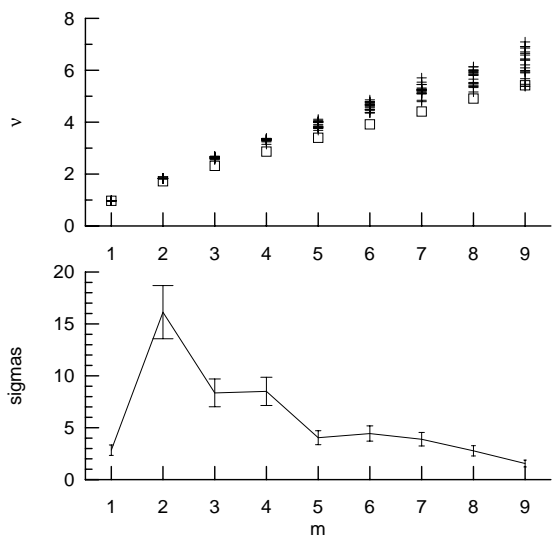

Figure 6: *Evidence for nonlinearity in a time series composed of a sine wave plus white noise. While there is no indication for low-dimensionality (there is too much noise to see that the underlying signal is one-dimensional), the estimated dimension is significantly different for the original data than for the surrogate data.*

Consider again a sine wave with additive measurement noise, but to keep the analysis simple, take the sampling rate to be exactly twice the frequency of the signal. The time series is produced by $x_t = s_t + n_t$, where the signal $s_t = (-1)^t S$ is alternating in sign while maintaining a constant amplitude $S$, and the noise $n_t$ is a white noise process of amplitude $\sigma$. That is,

$$x_t = (-1)^t S + \sigma e_t, \tag{11}$$

where the $e_t$'s are unit variance IID Gaussian random variables. The time series is linear, and can be produced by the ARMA model

$$x_t = -x_{t-1} + \sigma e_t + \sigma e_{t-1}, \tag{12}$$

with appropriate initial conditions. In fact, the initial conditions are crucial; notice that the signal amplitude $S$ does not even appear in Eq. (12). This is a general property of processes with infinite coherence times. For a process with a finite coherence time, the amplitude of the signal is determined solely by the coefficients of the ARMA model.[15]

---

[15]This is another disadvantage of ARMA surrogates compared to FT surrogates; the FT surrogates by construction will possess the same amplitude as the original time series.

### 3.3.1 Linear modeling

Let $M_o(m)$ denote the best order-$m$ autoregressive (AR) linear predictor; here

$$\hat{x}_t = \sum_{k=1}^{m} a_k x_{t-k}, \tag{13}$$

where the coefficients are chosen to minimize the mean squared error of the process:

$$\langle (x_t - \hat{x}_t)^2 \rangle = \langle (x_t - \sum_{k=1}^{m} a_k x_{t-k})^2 \rangle \tag{14}$$

$$= \left[ 1 - \sum_{k=1}^{m} (-1)^k a_k \right]^2 S^2 + \left[ 1 + \sum_{k=1}^{m} a_k^2 \right] \sigma^2. \tag{15}$$

With a little algebra, one can show that these coefficients are given by

$$a_k = (-1)^k \frac{S^2}{mS^2 + \sigma^2}, \tag{16}$$

and that the average squared error of this optimal predictor is given by

$$E^2[M_o(m)] = \langle (x_t - \hat{x}_t)^2 \rangle = \sigma^2 + \frac{\sigma^2 S^2}{mS^2 + \sigma^2}. \tag{17}$$

Note that the error decreases monotonically with increasing $m$, approaching a "floor" of $\sigma^2$ as $m \to \infty$. It is basically impossible to beat this error with any model, linear or nonlinear, because it is the noise on the signal which is by definition unpredictable. It is also worth remarking that the convergence of the linear AR model is algebraically slow with embedding dimension $m$. For chaotic processes (or more generally, for any stochastic processes with a finite coherence time), the convergence is usually faster.

This error assumes that the "correct" model is chosen for a given order $m$. In practice, one fits a model $M_s$ to a finite sample of $N$ data points. The fit is optimal for the data in the sample set, but in general is not optimal for out-of-sample data. Particularly when $m$ is large, and $N$ is small, the difference between the out-of-sample error for "correct" model and for the *fit* model can be significant.

The effect can be quantified with the aid of the Akaike Information Criterion (AIC) [1], which provides a measure of the difference between in-sample error $E_s$ and out-of-sample error $E_o$ for the best in-sample model $M_s$. (See Tong [53, §5.4] for a modern discussion.) Here,

$$\log(E_o[M_s(m, N)]) = \log(E_s[M_s(m, N)]) + m/N. \tag{18}$$

We have observed numerically that the error of the "correct" model $M_o$ lies roughly half way between the in-sample and out-of-sample error of the in-sample fit model $M_s$; that is, the difference $m/N$ in Eq. (18) can be split into two roughly equal components:[16]

$$\log(E_o[M_s(m, N)]) - \log(E_s[M_s(m, N)]) =$$

---

[16]S. Ellner (personal communication) has provided a heuristic argument for why the terms should be equal. The argument notes that the difference in error between $M_o$ and $M_s$ can be expanded as a Taylor expansion in $\theta_o - \theta_s$ (where $\theta$ represents the finite vector of parameters in model $M$), and that the relevant term is the second derivative of $E_o$ and $E_s$, respectively, multiplied by $(\theta_o - \theta_s) \cdot (\theta_o - \theta_s)$. Since one expects $E_s$ and $E_o$ to be asymptotically equal (as $N \to \infty$), it follows that their second derivatives should also be asymptotically equal; thus the expected differences $E_s[M_o] - E_s[M_s]$ and $E_o[M_s] - E_o[M_o]$ should also approach equality for large $N$.

$$\{\log(E_o[M_s(m, N)]) - \log(E_o[M_o(m)])\}$$
$$+ \quad \{\log(E_s[M_o(m)]) - \log(E_s[M_s(m, N)])\} \tag{19}$$

where

$$\log E_o[M_s(m, N)] - \log E_o[M_o(m)] \approx m/2N \tag{20}$$

and

$$\log E_o[M_o(m)] - \log E_s[M_s(m, N)] =$$
$$\log E_s[M_o(m)] - \log E_s[M_s(m, N)] \approx m/2N. \tag{21}$$

For large $N$ and large $m$ (but $m \ll N$), we can combine Eq. (17) and Eq. (21) to write the total squared error as the sum of noise (unavoidable), model inadequacy ($m$ too small), and parameter mis-specification ($N$ too small):

$$E^2[M_s(m, N)] = \sigma^2 + \frac{\sigma^2}{m} + \frac{m\sigma^2}{N} \tag{22}$$

Thus, for a given finite $N$, there will be an optimum $m$ for which the total error is minimized. In particular, for large $N$, the optimum model occurs when $m = \sqrt{N}$, and the total squared error in this case is given by $\sigma^2 + 2\sigma^2/\sqrt{N}$.

### 3.3.2 Nonlinear modeling

By contrast, consider as an example, the following parametric nonlinear model:

$$\hat{x}_t = -S^* \mathrm{sgn}(x_{t-1}) \tag{23}$$

where sgn is the "signum" or "sign" function; it's value is +1 or -1, depending on the sign of its argument. Here $m = 1$, and using a learning set of size $N$, one can estimate the parameter $S^*$ to within an error of $\sigma/\sqrt{N}$. (This assumes $\sigma \ll S$ so that $\mathrm{sgn}(x_t) = \mathrm{sgn}(s_t)$ at almost every time step.) Then, the total squared error is given by

$$\langle (x_t - \hat{x}_t)^2 \rangle = \langle (\sigma e_t + (S - S^*)\mathrm{sgn}(x_{t-1}))^2 \rangle \tag{24}$$
$$= \sigma^2 + \langle (S - S^*)^2 \rangle \tag{25}$$
$$= \sigma^2 + \sigma^2/N. \tag{26}$$

Though both the linear and nonlinear model converge to the same "floor" in the $N \to \infty$ limit, the nonlinear model converges more quickly. For a given $N$, the nonlinear model (with $m = 1$) beats the best linear AR model (with any $m$).

One might argue that using this parametric form for the nonlinear model is unfair, since in general one does not know the nature of the model that generated the time series. However, we remark that this model is not far from a local linear approximation that uses $N/2$ nearest neighbors.[17]

---

[17]If $\sigma$ is small, then the $N/2$ neighbors of a point with $x_t > 0$ will be a cloud of points which all have $x_t > 0$. A linear fit to this data will be of the form $\hat{x}_{t+1} = A + B(x_t - S)$ where $A \approx S$, in particular $A - S \sim \sigma/\sqrt{N}$, and $B \sim 1/\sqrt{N}$. It follows that the reduced squared error $\langle (S - \hat{x}_t)^2 \rangle$ will scale like $\sigma^2/N$. By contrast the linear $m = 1$ model, achieves $\langle (S - \hat{x}_t)^2 \rangle \sim \sigma^2$. Already, at $m = 1$, the local-linear model is better than the best global linear model, which requires an embedding dimension $m = \sqrt{N}$, and achieves a reduced squared error that scales as $\sigma^2/\sqrt{N}$.

In the example in Fig. 7, we consider $S = 1$, $\sigma = 0.3$, and $N = 128$. The ratio of signal to noise power is $S^2/\sigma^2 = 11$. Figure 7 shows both the theoretical error and the results of numerical simulations for these parameters.

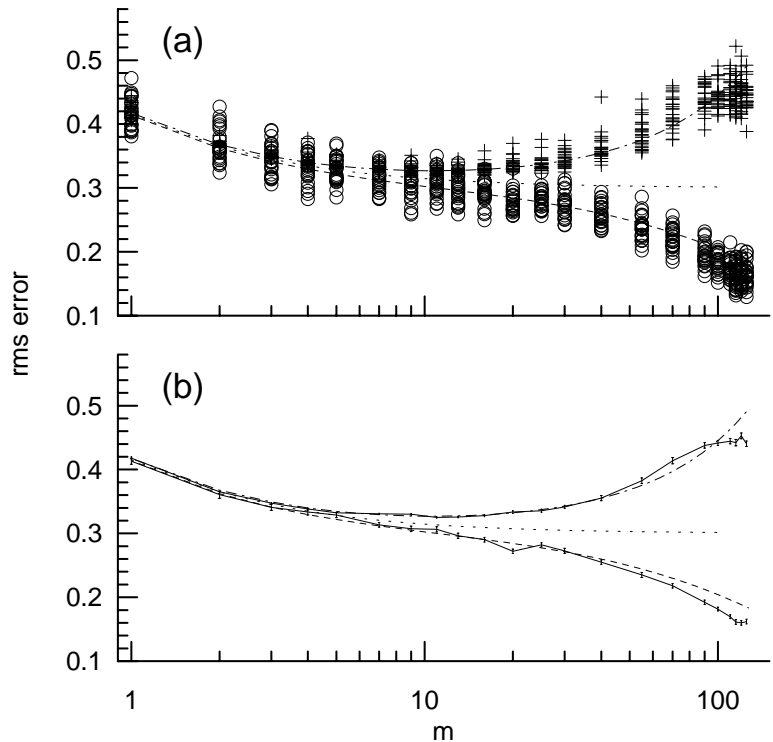

Figure 7: *In-sample and out-of-sample errors are plotted against embedding dimension, for linear AR models fit to $N = 128$ data points of a signal plus noise process defined in Eq. (11), with signal amplitude $S = 1$ and noise amplitude $\sigma = 0.3$. (a) Circles denote in-sample errors and pluses are out-of-sample errors for individual trials. (b) Median (with error bars given as the standard error) of the errors shown in the above panel. In both (a) and (b), we have plotted three theoretical curves. The dotted line corresponds to the expected error $E[M_o(m)]$ of the "correct" order-m model, given in Eq. (17). The dashed line is the theoretical in-sample error of the best-fit model $E_s[M_s(m, N)]$, given in Eq. (21), and the dashed-dotted line is the theoretical out-of-sample error of the best-fit model $E_o[M_s(m, N)]$, given in Eq. (20).*

### 3.4 Nonsinusoidal periodic signals

The difficulties associated with periodic sine signals seem to be compounded when higher harmonics are added. (Some of these extra difficulties were also addressed in Ref. [42].) Consider a time series given by $x(t) = \sin(t) + \sin(3t) + \sigma e_t$, where $\sigma = 0.05$ and $e_t$ is uniform noise with unit range. Even for very small $\sigma$, a linear model of this data requires four past values to predict the future, because it has to estimate the phase and amplitude of two sine waves — the phase and amplitude are *not* coded into the model itself. One finds that for the same embedding dimension $m$, nonlinear models fit this data better than linear models.

In particular, as seen in Fig. 8, Casdagli's plots of forecasting error as a function of number of neighbors in the local linear fit [8, 9] indicate nonlinearity in a time series, even though the system is formally speaking linear. Our intuitive explanation is that the nonlinear models are able to use information that is unavailable to the direct linear model; namely the amplitudes and relative phase of the two sine waves. So, while the linear model requires four degrees of freedom, the nonlinear model is relatively successful with only one. We should emphasize that the Casdagli plot was intended as an "exploratory" method of time series analysis, and that it appears very well suited for that purpose. The ambiguity of interpretation that arises when the Casdagli plot is applied to data with long coherence times is a problem that is not unique to the Casdagli plot, but is just another artifact of the long coherence time.[18]

And in generating surrogate data sets, one again finds that nonsinusoidal periodicity is even worse than sinusoidal. As well as the usual difficulties, one has the added problem that the FT algorithm does not preserve the phase relation between the harmonics. It is this phase relation that determines the shape of the periodic waveform. ARMA modeling is even worse, because in that case, the model encodes neither the phase relation between the harmonics *nor* the relative amplitudes of the sinusoidal components.

### 3.5 Aside: Chaos and long coherence times

Although the situation we have described so far has been restricted to linear systems with noise, we note that fully deterministic chaotic systems can also exhibit long coherence times. While this may seem at first counter-intuitive, since positive Lyapunov exponents imply a finite "forgetting" time, the effect has been previously noted [13, 14] in the context of the Rössler flow [41], and is readily apparent in maps which exhibit "banded chaos." An example of the latter is the logistic map, $x_{t+1} = \lambda x_t(1 - x_t)$, at parameter $\lambda = 3.6$. The attractor is chaotic, but the orbit alternately visits two bands, one above and one below the fixed point at $x = 0.72$. This underlying period two motion is coherent over the full length of the trajectory.

## 4 Discussion

We provide three possible interpretations of the basic source of the problems that arise when surrogates of highly coherent time series are generated. The first is technical; the second and third have more of a philosophical, almost existential flavor.

### 4.1 The surrogate data generator is flawed.

One might argue that the inability of the FT algorithm to generate surrogates which mimic the original data indicates a flaw in the algorithm. For coherent signals, the true power spectrum contains instrumentally sharp spikes. However, when estimating the power spectrum from a finite time series, the spike is spread out over several distinct frequency bins with a very specific phase relation between them. When these phases are scrambled, and the FT is inverted, the resulting time series has a shorter coherence time.

One can imagine various *ad hoc* solutions, such as randomizing phases only for frequencies not in the vicinity of the dominant frequency. We have not investigated such modifications, and are hesitant to do so, since they are difficult to automate in a way that would be applicable to all time series. (For example, suppose one has a quasiperiodic time series, and that each of the component frequencies also has higher harmonic frequencies. Distinguishing the peaks from the broadband then becomes a nontrivial task.)

---

[18] We are tempted to say that the problem lies not in the analysis but in the data itself!

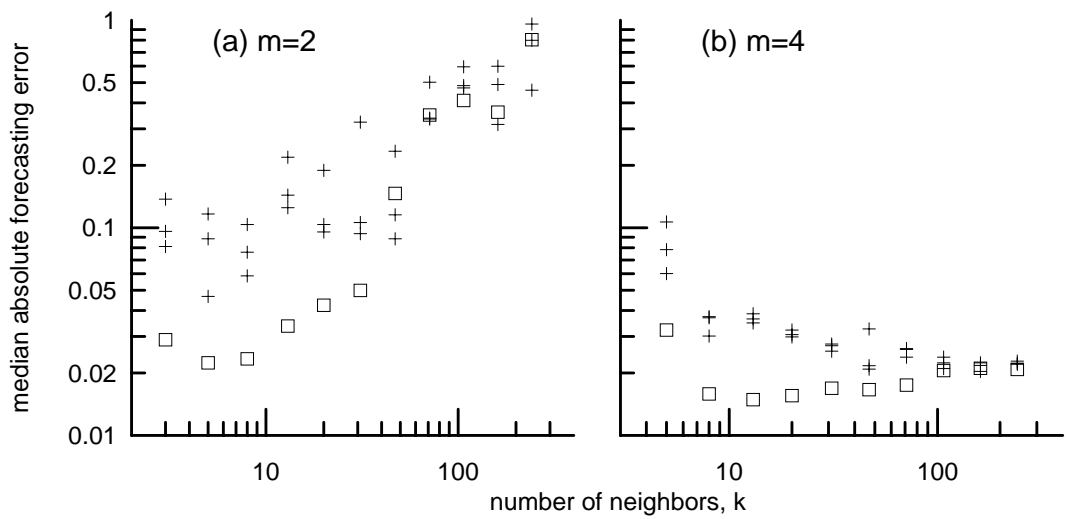

Figure 8: *Plot in the style of Casdagli for a time series generated by adding two sine waves and a small measure of white noise (□), as well as for some surrogate time series (+). Although this is by construction a linear time series, the plot of forecasting error versus number of neighbors in the local linear predictor indicates that nonlinear models are superior to linear models. Here, $N = 1024$ points are taken from the time series, and the embedding dimension is (a) $m = 2$, and (b) $m = 4$, which is in principle adequate for a linear model of two sine waves. Note that comparison with surrogate data also implies that the time series is nonlinear.*

Technical problems arise with ARMA models as well; namely with stability, and the difficulty of choosing the coefficients "just right." For ARMA modeling of the sine wave, even if the correct coefficients are chosen, there is no way to assure that the surrogates will be the same amplitude as the original time series; the amplitude of linear models is coded not in the coefficients but in the initial conditions. One must not only model the coefficients of a linear model, then, one must further restrict the initial conditions which are iterated. (Normally, when the coherence time is finite, the amplitude *is* specified by the coefficients, though its dependence becomes increasingly sensitive as the coherence time increases.) Rescaling so that the amplitude of the surrogates matches that of the original time series does not really solve this problem, because if the signal is composed of several sine waves, one must also find a way to maintain all their *relative* amplitudes.

An interesting possibility which we have not pursued is to model the time series not as an ARMA but directly in terms of its deterministic and nondeterministic components; *i.e.,* Eqs. (1-3). The surrogates would be generated with new white noise realizations in Eq. (3) for the indeterministic component, but the deterministic (coherent) component would be kept the same as the original.

## 4.2 The time series is nonstationary.

If the fault is not in the algorithm, then perhaps it is in the data. While *stationarity* has a clear-cut meaning for a stochastic *process*, it is a fuzzier concept when applied to a *time series*. The Lorenz flow [32] is a stationary chaotic process, but if a time series is taken over a short enough segment, it will appear very nonstationary. For a time series, we argue that an useful operational definition of *stationarity* is that the characteristic time scales in the data are much shorter than than the length of the data set itself.

If we think of the coherence time a one of the characteristic time scales, then highly coherent time series are not stationary.

It may seem odd to characterize a sinusoidal signal as nonstationary. But one way to see why this is reasonable is to consider two sine waves whose frequencies are nearly equal. The sum of the two sine waves will exhibit a low frequency beating as they slowly move in and out of phase with each other. If the length of the time series is shorter than the beating period, then the resulting time series will appear quite nonstationary. So, if we'd like the sum of two stationary time series to itself be a stationary time series, we cannot permit time series with long coherence times to be considered stationary.

## 4.3 The time series is nonlinear.

A third interpretation of the spurious identification of nonlinearity in time series with long coherence times is that the nonlinearity is not spurious at all. Typical linear processes do not produce long coherence times, because their parameters need to be precisely adjusted.[19]

---

[19]For example, to generate a sine wave with additive noise (see Section 3.2) requires an ARMA(2,2) model $x_t = a_1 x_{t-1} + a_2 x_{t-2} + \sigma e_t + b_1 e_{t-1} + b_2 e_{t-2}$, where the roots of $z^2 = a_1 z + a_2$ must lie precisely on the unit circle ($|a_1| < 2$, $a_2 = -1$; or $a_1 = -1$, $a_2 = 0$), and $b_1$ and $b_2$ must be precisely equal to $-\sigma a_1$ and $-\sigma a_2$, respectively. If the roots are outside the unit circle, then the time series will diverge to infinity exponentially;

We note that those clean and highly coherent sine waves that come out of signal generators in the laboratory do not arise from RLC circuits, but depend crucially on the nonlinearity of the electronics. In nature, and in the laboratory, nonlinear limit cycles are very common and very robust. Thus, one might argue that a long coherence time is in itself evidence for underlying nonlinear dynamics.

## 4.4  Summary

Although a time series may be generated by a process that is formally linear, if it has a long coherence time it can often fool tests for linearity, and can be mistaken for nonlinear time series. In particular, it is difficult to generate surrogate data which mimics the linear properties of the process that generated the data.

In testing a time series for nonlinearity, it is a good idea to compare with surrogate data using both nonlinear *and* linear statistics. Good evidence for nonlinearity requires that the nonlinear statistics do distinguish the real and surrogate data, and that the linear statistics do not. Even more important is to plot an autocorrelation curve for the real data, and make sure that the autocorrelation $A(T)$ vanishes as $T$ gets large. If there is significant autocorrelation for $T$ on the order of the length of the time series, then one must beware the dangers of long coherence times.

Regarding E.dat, we did see some evidence for nonlinearity, but we note that that evidence is seen only for the smoothed data, and that it is "significant" but not "substantial." Finally, because E.dat has a long coherence time, we are further inclined to discount this evidence for nonlinearity.

## Acknowledgements

JT acknowledges many useful conversations with Dean Prichard, Steve Ellner, and the LANL/SFI Half Baked Lunch Time NonLinear Time Series Working Group. We are also pleased to express appreciation to Blake LeBaron and many other conference participants whose comments and suggestions have diffused into this manuscript, in one form or another, usually without attribution. Thanks also to Lenny Smith and Andreas Weigend for a critical reading of the manuscript.

Work by JT was partially supported by National Institute of Mental Health grant 1-R01-MH47184, and performed under the auspices of the Department of Energy. PSL thanks the Office of Naval Research and the Department of Energy for support. DMR was supported by NASA grant W-17,975.

Finally, we want to thank Neil Gershenfeld and Andreas Weigend for inviting us (JT and PSL) to this unique and fascinating conference, even though we lacked the discipline (or was it the courage?) to actually submit an entry to the contest.

---

and if the roots are precisely on the unit circle but $b_i \neq -\sigma a_i$, the time series will diverge to infinity like a random walk. If the roots are inside the unit circle, then there will be a finite coherence time.

# References

1. H. Akaike, "A new look at the statistical model identification," *IEEE Trans. Auto. Control* **19**, 716–723 (1974).

2. T. W. Anderson, *The statistical analysis of time series.* (Wiley, New York, 1971).

3. W. A. Brock and W. D. Dechert, "Statistical inference theory for measures of complexity in chaos theory and nonlinear science," in *Measures of Complexity and Chaos*, N. Abraham *et al.*, eds. (Plenum, 1989), pp. 79–98.

4. W. A. Brock, W. D. Dechert, and J. Scheinkman, "A test for independence based on the correlation dimension," Technical Report 8702, Social Systems Research Institute, University of Wisconsin, Madison (1987).

5. W. A. Brock, J. Lakonishok, and B. LeBaron, "Simple technical trading rules and the stochastic properties of stock returns," *J. Finance* **47**, 1731–1764 (1992).

6. W. A. Brock and S. M. Potter, "Diagnostic testing for nonlinearity, chaos, and general dependence in time series data," in *Nonlinear Modeling and Forecasting*, M. Casdagli and S. Eubank, eds. vol. XII of *SFI Studies in the Sciences of Complexity*, (Addison-Wesley, 1992), pp. 137–162.

7. W. A. Brock and C. L. Sayers, "Is the business cycle characterized by deterministic chaos?," *J. Monetary Econ.* **22**, 71–90 (1988).

8. M. Casdagli, "Chaos and deterministic versus stochastic nonlinear modeling," *J. R. Stat. Soc. B* **54**, 303–328 (1992).

9. M. Casdagli, "Nonlinear forecasting, chaos and statistics," in *Modeling Complex Phenomena*, L. Lam and V. Naroditsky, eds. (Springer-Verlag, New York, 1992), pp. 131–152.

10. B. Efron, "Computers and the theory of statistics: thinking the unthinkable," *SIAM Review* **21**, 460–480 (1979).

11. B. Efron and R. Tsibirani, "Bootstrap methods for standard errors, confidence intervals, and other measures of statistical accuracy," *Statistical Science* **1**, 54–77 (1986).

12. S. Ellner, "Detecting low-dimensional chaos in population dynamics data: a critical review," in *Chaos and Insect Ecology*, J. A. Logan and F. P. Hain, eds. (University of Virginia Press, Blacksburg, VA, 1991), pp. 65–92.

13. D. Farmer, J. Crutchfield, H. Froehling, N. Packard, and R. Shaw, "Power spectra and mixing properties of strange attractors," *Annals of the New York Academy of Science* **357**, 453–472 (1980).

14. J. D. Farmer, "Spectral broadening of period-doubling bifurcation sequences," *Phys. Rev. Lett.* **47**, 179–182 (1981).

15. P. Grassberger, "Do climatic attractors exist?," *Nature* **323**, 609–612 (1986).

16. P. Grassberger and I. Procaccia, "Characterization of strange attractors," *Phys. Rev. Lett.* **50**, 346–349 (1983).

17. P. Grassberger and I. Procaccia, "Measuring the strangeness of strange attractors," *Physica D* **9**, 189–208 (1983).

18. M. J. Hinich, "Testing for gaussianity and linearity of a stationary time series," *J. Time Series Anal.* **3**, 169–176 (1982).

19. D. A. Hsieh, "Testing for nonlinear dependence in daily foreign exchange rate changes," *J. Business* **62**, 339–368 (1989).

20. D. A. Hsieh, "Chaos and nonlinear dynamics: application to financial markets," *J. Finance* **46**, 1839–1877 (1991).

21. G. M. Jenkins and D. G. Watts, *Spectral analysis and its applications.* (Holden-Day, San Francisco, 1968).

22. M. Kanter, "Lower bounds for nonlinear prediction error in moving average processes," *Ann. Prob.* **7**, 128–138 (1979).

23. D. T. Kaplan and R. J. Cohen, "Is fibrillation chaos?," *Circulation Res.* **67**, 886–892 (1990).

24. D. M. Keenan, "A Tukey nonadditivity-type test for time series nonlinearity," *Biometrika* **72**, 39–44 (1985).

25. M. B. Kennel and S. Isabelle, "Method to distinguish possible chaos from colored noise and to determine embedding parameters," *Phys. Rev. A* **46**, 3111–3118 (1992).

26. H. R. Künsch, "The jackknife and the bootstrap for general stationary observations," *Ann. Stat.* **17**, 1217–1241 (1989).

27. J. Kurths and H. Herzel, "An attractor in a solar time series," *Physica D* **25**, 165–172 (1987).

28. B. LeBaron, "Do moving average trading rule results imply nonlinearities in foreign exchange markets?," Technical Report 9222, Social Systems Research Institute, University of Wisconsin, Madison (1992).

29. T.-H. Lee, H. White, and C. W. J. Granger, "Testing for neglected nonlinearity in time series models: A comparison of neural network methods and alternative tests." To appear in *J. Econometrics*.

30. C. Leger, D. N. Politis, and J. P. Romano, "Bootstrap technology and applications." To appear in *Technometrics* (1992).

31. K. S. Lii and M. Rosenblatt, "Deconvolution and estimation of transfer function phase and coefficients for nongaussian linear processes," *Ann. Stat.* **10**, 1195–1208 (1982).

32. E. N. Lorenz, "Deterministic nonperiodic flow," *J. Atmos. Sci.* **20**, 130–141 (1963).

33. A. I. McLeod and W. K. Li, "Diagnostic checking ARMA time series models using squared-residual autocorrelations," *J. Time Series Anal.* **4**, 269–273 (1983).

34. J. E. Moody, "The effective number of parameters: an analysis of generalization and regularization in nonlinear learning systems," in *Advances in Neural Information Processing Systems 4*, J. E. Moody, S. J. Hanson, and R. P. Lippmann, eds. (Morgan Kaufmann, San Mateo, CA, 1992).

35. A. R. Osborne, A. D. Kirwin, A. Provenzale, and L. Bergamasco, "A search for chaotic behavior in large and mesoscale motions in the Pacific Ocean," *Physica D* **23**, 75–83 (1986).

36. A. R. Osborne and A. Provenzale, "Finite correlation dimension for stochastic systems with power-law spectra," *Physica D* **35**, 357–381 (1989).

37. D. N. Politis and J. P. Romano, "The stationary bootstrap," Technical Report 365, Department of Statistics, Stanford University (1991).

38. W. H. Press, B. P. Flannery, S. A. Teukolsky, and W. T. Vetterling, *Numerical Recipes.* (Cambridge University Press, Cambridge, 1988).

39. A. Provenzale, A. R. Osborne, and R. Soj, "Convergence of the $K_2$ entropy for random noises with power law spectra," *Physica D* **47**, 361–372 (1991).

40. T. Pukkila, S. Koreisha, and A. Kallinen, "The identification of ARMA models," *Biometrika* **77**, 537–548 (1990).

41. O. E. Rössler, "An equation for continuous chaos," *Phys. Lett. A* **57**, 397–398 (1976).

42. D. M. Rubin, "Use of forecasting signatures to help distinguish periodicity, randomness, and chaos in ripples and other spatial patterns," *Chaos* **2**, 525–535 (1992).

43. J. A. Scheinkman and B. LeBaron, "Nonlinear dynamics and stock returns," *J. Business* **62**, 311–338 (1989).

44. G. Schwarz, "Estimating the dimension of a model," *Ann. Stat.* **6**, 461–464 (1978).

45. L. A. Smith, "Identification and prediction of low dimensional dynamics," *Physica D* **58**, 50–76 (1992).

46. T. Subba Rao and M. M. Gabr, "A test for linearity of stationary time series," *J. Time Series*

*Anal.* **1**, 145–158 (1980).

47. J. Theiler, *Quantifying Chaos: Practical Estimation of the Correlation Dimension.* PhD thesis, Caltech (1988).

48. J. Theiler, "Estimating fractal dimension," *J. Opt. Soc. Am. A* **7**, 1055–1073 (1990).

49. J. Theiler, "Some comments on the correlation dimension of $1/f^\alpha$ noise," *Phys. Lett. A* **155**, 480–493 (1991).

50. J. Theiler and S. Eubank, "Don't bleach chaotic data," Technical Report LA-UR-92-1575, Los Alamos (1992).

51. J. Theiler, S. Eubank, A. Longtin, B. Galdrikian, and J. D. Farmer, "Testing for nonlinearity in time series: the method of surrogate data," *Physica D* **58**, 77–94 (1992).

52. J. Theiler, B. Galdrikian, A. Longtin, S. Eubank, and J. D. Farmer, "Using surrogate data to detect nonlinearity in time series," in *Nonlinear Modeling and Forecasting*, M. Casdagli and S. Eubank, eds. vol. XII of *SFI Studies in the Sciences of Complexity*, (Addison-Wesley, 1992), pp. 163–188.

53. H. Tong, *Non-linear Time Series: A Dynamical System Approach.* (Clarendon Press, Oxford, 1990).

54. R. S. Tsay, "Nonlinearity tests for time series," *Biometrika* **73**, 461–466 (1986).

55. R. S. Tsay, "Nonlinear time series analysis: diagnostics and modelling," *Statistical Sinica* **1**, 431–451 (1991).

56. R. S. Tsay, "Model checking via parametric bootstraps in time series analysis," *Applied Statistics* **41**, 1–15 (1992).

57. A. Weigend, B. Huberman, and D. Rummelhart, "Predicting the future: a connectionist approach," *Int. J. Neural Systems* **1**, 193–206 (1990).